\documentstyle[prl,aps,twocolumn]{revtex}

\begin{document}
\preprint{APS/123-QED}
\wideabs{
\title{Anharmonic vibrations in nuclei}
\author{M. Fallot$^{a)}$, Ph. Chomaz$^{b)}$, M.V. Andr\'{e}s$^{c)}$, F. Catara$^{d)}$%
, E. G. Lanza$^{d)}$, J. A. Scarpaci$^{a)}$}
\address{$^{a)}$ Institut de Physique Nucl\'{e}aire, IN2P3-CNRS, F-91406 Orsay Cedex,
France \\
$^{b)}$ GANIL, B.P. 5027, F-14076 CAEN Cedex 5, France \\
$^{c)}$ Departamento de F\'{\i }sica At\'{o}mica, Molecular y Nuclear,
Universidad de Sevilla, Apdo 1065, E-41080 Sevilla, Spain \\
$^{d)}$ Dipartimento di Fisica Universit\'{a} di Catania and INFN, Sezione
di Catania, I-95129 Catania, Italy}

\maketitle
\begin{abstract}
In this letter, we show that the non-linearities of large amplitude
motions in atomic nuclei induce giant quadrupole and monopole
vibrations. As a consequence, the main source of anharmonicity is the
coupling with configurations including one of these two giant
resonances on top of any state. Two-phonon energies are often lowered
by one or two MeV because of the large matrix elements with
such three phonon configurations. These effects are studied in two
nuclei, $^{40}$Ca and $^{208}$Pb.
%\pacs{21.60Ev, 21.10Re, 21.60Jz, 24.30Cz}% PACS, the Physics and Astronomy
\end{abstract}
}
\smallskip Many-body fermionic systems possess collective vibrational states
which are well described as bosonic modes (phonons). The existence in atomic
nuclei of such states, both low-lying and Giant
Resonances (GR) is now well established up to the second quantum \cite
{grbook,report}. However, their properties such as
energy and excitation probability are still open questions. From the
experimental point of view the strong excitation cross section of two phonon
states calls for the presence of large anharmonicities { but up to
now, all the theoretical estimates 
were pointing to weak deviations from a harmonic spectrum.} 
To our knowledge, so far only the mixing of one- and two-phonon states has
been considered in microscopic calculations, with two exceptions. In ref.~%
\cite{popo1} the coupling to some specific three-phonon configurations has
been included as a mechanism generating the damping width of the
Double Giant Dipole Resonance. In ref.~\cite{popo2} the fragmentation
of the doubly excited low lying octupole states in $^{208}$Pb
has been studied by allowing the coupling to one- and three-phonon
configurations with a low energy cut-off introduced to reduce the
diagonalization space. For this reason monopole, (GMR) and quadrupole
(GQR) contributions which, as we will see, play an important role, were
neglected.

In the present paper we show that a correct description of the states { for which
the} main
component is a two-phonon configuration requires the inclusion of one- and
three-phonon ones. We stress the essential role played by the breathing mode
in the nuclear anharmonicity as an important novelty of the present analysis
since volume modes are usually not considered in damping or coupling
mechanisms. Moreover, we will show that the very collective GQR plays also
an important role. 

The starting point of our calculation is a mapping of the fermion
particle-hole operators $a_{p}^{\dagger }a_{h}$ into boson operators $%
B_{ph}^{\dagger }$ as for example the one proposed in ref.~\cite{lamb}
\begin{eqnarray}
a_{p}^{\dagger }a_{h} &\rightarrow &B_{ph}^{\dagger }+(1-\sqrt{2}%
)\sum_{p^{\prime }h^{\prime }}B_{p^{\prime }h^{\prime }}^{\dagger
}B_{p^{\prime }h}^{\dagger }B_{ph^{\prime }}+....  \label{aa} \\
a_{p}^{\dagger }a_{p^{\prime }} &\rightarrow &\sum_{h}B_{ph}^{\dagger
}B_{p^{\prime }h}\;,\;\;a_{h}a_{h^{\prime }}^{\dagger }\rightarrow
\sum_{p}B_{ph}^{\dagger }B_{ph^{\prime }}  \label{bb}
\end{eqnarray}
where $a^{\dagger }$ ($a$) creates (annihilates) one nucleon in an occupied (%
$h$) or unoccupied ($p$) single particle state. The second term on the right
hand side of eq.(\ref{aa}) is a correction that takes care of the Pauli
principle. Then we construct a boson image of the Hamiltonian, truncated 
at the fourth order in the $%
B^{\dagger }$ and $B$ operators.
Introducing the Bogoliubov transformation for bosons: 
\begin{equation}
Q_{\nu }^{\dagger }\equiv \sum_{p,h}(X_{ph}^{\nu }B_{ph}^{\dagger
}-Y_{ph}^{\nu }B_{ph})  \label{bogo}
\end{equation}
and imposing that the quadratic part of the boson Hamiltonian 
in the new operators is diagonal, we obtain the usual { Random Phase 
Approximation (RPA)} equations for the $X$ and 
$Y$ amplitudes. 

By inverting eq.(\ref{bogo}) we can express $H_{B}$ in terms of the
collective $Q^{\dagger }$ and $Q$ operators: 
\begin{eqnarray}
H_{B} &=&{\cal H}_{11}Q^{\dagger }Q+({\cal H}_{21}Q^{\dagger }Q^{\dagger
}Q+h.c.)+{\cal H}_{22}Q^{\dagger }Q^{\dagger }QQ  \nonumber\\
&&+({\cal H}_{30}Q^{\dagger }Q^{\dagger }Q^{\dagger }+h.c.)+({\cal H}%
_{31}Q^{\dagger }Q^{\dagger }Q^{\dagger }Q+h.c.)  \nonumber \\
&&+({\cal H}_{40}Q^{\dagger }Q^{\dagger }Q^{\dagger }Q^{\dagger }+h.c.)+\ldots  
\label{hquart} 
\end{eqnarray}
with ${\cal H}_{11}^{\nu \nu ^{\prime }}=E_{\nu }\delta _{\nu \nu
^{\prime }}.$ The ${\cal H}$ matrices are expressed in terms of the X and Y
of transformation (\ref{bogo}). The contributions to eq.(\ref{hquart})
coming from the high order terms of the expansion (\ref{aa}) appear to
be reduced by the number of configurations involved in the collective
states and therefore can be neglected. In the case of closed shell
nuclei, the RPA correlations are moderate. Therefore, the Y/X ratios
are small. In eq.(\ref{hquart}) we will neglect the ${\cal H}$ terms
containing at least one Y amplitude. 
These two approximations leave unaffected only the
first three terms of eq.(\ref{hquart})
\cite{ccg2}. We will compare the spectra of
$^{40}$Ca and $^{208}$Pb obtained by the
diagonalization in the spaces
containing up to two-phonon states and up to three-phonon states,
respectively. In ref. \cite{lip} a similar analysis was done in the
two level Lipkin model. It was found that this approximation is well
justified and one gets good results in the larger space for the
eigenstates { which} main component is a two-phonon configuration.

All calculations have been performed by using the SGII Skyrme
interaction~\cite{sg}. We include all natural parity RPA collective
one-phonon states with angular momentum $J\le 3$ which exhaust at
least $5\%$ of the EWSR and all two- and three-phonon configurations
built with them, without any energy cut-off, with both natural and
unnatural parity.

Let us start looking at the results for $^{40}$Ca. In table \ref{ca40.rpa}
we show the one-phonon states taken into account. { In table 
\ref{ca40.res} we show some results of the diagonalization for a selected
set of states.}
The energies obtained in the space up to two-phonons { (see ref. \cite{lanza1})}
are reported
here for comparison. The so-calculated anharmonicity was limited to a few hundred
keV.
Let us now study the more complete calculation including the three
phonon states.

As a general comment, one can say that the shift induced by the coupling to
three-phonon states is fairly large, being in almost all { the} cases more than 1
MeV, and always downward. This can be understood in second order
perturbation which, as can be seen from the table, gives a good estimate
for the energies in most cases. In second order perturbation the correction
to the energy is given by 
\begin{equation}
\Delta E_{i}=<\varphi _{i}|V|\varphi _{i}>+\sum_{j\ne i}{\frac{{|<\varphi
_{j}|V|\varphi _{i}>|^{2}}}{{E_{i}^{0}-E_{j}^{0}}}}  \label{pert}
\end{equation}
where $|\varphi _{i}>$ is the considered unperturbed state, $|\varphi _{j}>$
{ all the other states} and $E^{0}$ the corresponding unperturbed energies. Since the
diagonal, first order, contribution is small in most cases, the sign of the
shift is that of the denominator in the second order term. Therefore, if $%
|\varphi _{i}>$ is a two-phonon state, the contributions from three-phonon
configurations are negative in most cases since most of the three phonon
states lye above the two phonon ones. Moreover, whenever a
GMR is added on top of any state, the ${\cal H}_{21}$
terms (see eq.\ref{hquart}) are large {, of the order of 1 to 2 
MeV in $^{40}$Ca. }
The specific values can be found in the last
three columns of table \ref
{ca40.rpa}.
 Indeed, in the coupling leading to the addition of one GMR on
top of any state, the residual interaction between the underlying
fermions is not truncated by conservation laws, because the particles
and the holes involved in the GMR carry identical parity and
spin quantum numbers (cf. ref. \cite{Beaumel}).
Phenomenologically, this strong coupling of all collective vibrations with the breathing
mode comes from the fact that in a small nucleus like the $^{40}$Ca any large
amplitude motion affects the central density. Therefore, surface modes
cannot be decoupled from a density variation in the whole volume as clearly
seen in recent TDHF\ simulations ref. \cite{Simenel}. 

If the state is a two-phonon one, then the matrix elements coupling it to
the state obtained by exciting a breathing mode on top of it
are about 3 MeV (up to 5.5 MeV)
when the less (more) collective component of the $^{40}$Ca GMR is
considered. Even larger matrix elements are obtained, when the states
connected by ${\cal H}_{21}$ involve several GMR. In that case a Bose
enhancement factor appears and no Clebsch-Gordan coefficients enter in the
calculation. Thus
the matrix element between the double and the triple GMR located at 18.25
MeV, $M_{1}$, is $\sqrt{6}$
times larger than between the single and the double $M_{1}$. That gives a
matrix element of -5.22 MeV, giving a contribution of -1.49 MeV to the
second order energy correction of the double $M_{1}$ state. An even larger
value comes out in the case of the double { GMR located at 22.47 MeV}, $M_{2}$, and the triple $M_{2}$,
namely a matrix element of -9.69 MeV giving a -4.18 MeV contribution to the
energy shift of the double $M_{2}$. This is due to the fact that $M_{2}$ is
more collective than $M_{1}$ in $^{40}$Ca. 

Something similar, but less strong, happens also for the matrix elements
connecting some state with that built by adding one GQR phonon. 
We quote two examples. The low-lying
component of the Giant Dipole Resonance $|D_{1}>$ has a matrix element of
the residual interaction with the states $|D_{1}$$~\otimes $$~M_{1}>$, $%
|D_{1}$$~\otimes $$~M_{2}>$ and $|D_{1}$$~\otimes $$~Q_{1}>$ equal to -1.38
MeV, -2.12 MeV and -1.25 MeV respectively. Another example, with total $J$%
=1, is given by the matrix elements between $|D_{1}$$~\otimes $$~Q_{1}>$ and $|(M_{1}$$%
~\otimes $$~D_{1})_{1}$$~\otimes $$~Q_{1}>$, $|(M_{2}$$~\otimes $$~D_{1})_{1}$%
$~\otimes $$~Q_{1}>$ and $|(Q_{1})_{2}^{2}$$~\otimes $$~D_{1}>$ equal to
-2.74 MeV, -4.61 MeV and -1.41 MeV respectively. 

These findings clearly indicate that large amplitude motions are strongly
coupled both to surface and volume oscillations, the latter being more
important in $^{40}$Ca. It is worthwhile stressing that such large
corrections to the energy of two-phonon states are obtained despite the quite large absolute values of 
the energy difference between the coupled states. 
Therefore, introducing an energy cut-off in the three-phonon
states included in the calculation may lead to erroneous results. Let us
consider for example the case of the $0^{+}$ member of the multiplet of
double low-lying octupole states. At first order perturbation, it is shifted
up by 2.24 MeV. The second order correction coming from the single GMR
states is -0.93 MeV. These two contributions, leading to a total shift of
+1.31 MeV, dominate the effects of the coupling with one- and two-phonon
states as confirmed by the result of the diagonalization in this subspace.
When three-phonon states are included, one gets a further shift down of 1.86
MeV coming from the configuration including a GMR on top of the two
octupoles. This contribution is absent in ref. \cite{popo2} because the energy
cut-off introduced there in order to reduce the number of three-phonon
configurations was too low. The same happens for the other members of the
multiplet as well as for the double $D_{1}$ or $D_{2}$, the double $Q_{1}$
and the $D_{1}$or $D_{2}$ $\otimes $ $Q_{1}$ states.

The results for $^{208}$Pb are shown in tables \ref{pb.rpa} and
\ref{pb.res}. The same general remarks already made for $^{40}$Ca
apply also in this case. The most relevant difference is that the role
played by the GMR and the GQR in $^{40}$Ca is now inverted, the latter
being dominant in $^{208}$Pb.  This reduced importance of the GMR
may come from the fact that in large nuclei the
surface vibrations can occur without changing the volume.
 Concluding about the energy of the two-phonon states
one can see that the inclusion of the three phonon configurations
induces an anharmonicity of more than 1 MeV in $^{40}$Ca but only of a
few hundred keV in $^{208}$Pb. Because of the location at high energy
of the three phonon states, the observed shift is systematically
downward. It is important to stress that 
the
considered residual interaction only couples states with a number of
phonon varying at maximum by one unit. 
{ Therefore, the energy variation of the two-phonon spectrum
induced by inclusion of four and more phonon states 
would be small since it corresponds to a third order perturbation 
involving two large energy
differences in the denominator.}

If we now analyze the splitting of the two-phonon multiplets we can see that
it remains small for giant resonances 
(about a few 
hundred keV)
while it may go up to 1 MeV for low lying states in $^{40}$Ca. Comparing the
splitting and the { ordering of the} states obtained in first order perturbation
and in the full calculation we can see that they remain almost unchanged.
Therefore, the diagonal matrix elements of the residual interaction are
responsible for this splitting and ordering. 

Let us now investigate the mixing induced by the residual
interaction. In tables \ref{ca40.res} and \ref{pb.res} the mixing
coefficients of the two main components { in each state} are presented. First we can
see that there is always one component that remains very large, explaining the
success of the perturbation approach. The important point is that in
general we observe large mixing coefficients, namely about $0.2$ to
$0.4$ or more in $^{40}$Ca and $0.15$ to $0.3$ in $^{208}$Pb. This may
have very important consequences in the excitation process as we will
investigate in a forthcoming work.

It is worthwhile mentioning that, in some cases, a three-phonon
component appears with a large amplitude in the wave function of a
(mainly) two-phonon state, despite the fact that the residual
interaction does not couple directly these configurations together. In
a few cases, indeed, this is the second main component as can be seen
for $^{40}$Ca in table \ref{ca40.res} (the $|(D_{1})_{0}^{2}>$ and
$|M_{2}$$~\otimes $$~Q_{1}>$ states) and for $^{208}$Pb in table \ref
{pb.res} (the $|(M_{1})^{2}>$ state). This happens because the
diagonal matrix elements of the Hamiltonian in the two-phonon and
three-phonon configurations are close and the matrix elements coupling
the latter with other configurations are large. A similar situation
has been found in $^{208} $Pb for two one-phonon (mainly) states which
have a three-phonon configuration as second important component, even
though our Hamiltonian does not couple directly states { which} numbers
of phonons differ by more than one.  This is the case for the state
{ which} main component is $|M_{1}>$, with amplitude $c_{0}$$=-0.79$,
{ and for which the second most} important component
is $|(3^{-})_{2}^{2}~\otimes2^{+}>$ 
with $c_{1}$$=0.55$. Both components have large matrix
elements with the two-phonon state $|(3^{-})_{0}^{2}>$.  How this
mixing of the monopole resonance may affect the monopole response, and
so the usual conclusion about the compressibility, is now under study.
The other case is the single high energy octupole resonance $|O>$
which is strongly mixed with the states 
$|(2^{+}$$~\otimes$$3^{-})_{J}$$~\otimes $$ Q_{1}>$. 
The energy of these
states are, however, shifted by less than 100 keV. This is coherent
because the strong mixing is coming from a quasi degeneracy of the
considered states.

Summarizing, the spectrum of two-phonon states is strongly modified by
their coupling to the three-phonon ones. All of the states appear
mixed with the excitation of a GMR and GQR on top of it. This is due
to the fact that most of the matrix elements of ${\cal H}_{21}$ 
coupling a phonon with the same phonon plus a GMR or a GQR are
large. Moreover, because of the Bose enhancement factors, the effect of
${\cal H}_{21}$ between the two and three phonon states is even
larger. It is also to be noted that many of the important three-phonon
states are higher in energy than the two-phonon ones.  Therefore, they
induce a systematic shift down of the two phonon states as the sum of
several quite large negative contributions. This
unexpected finding can be understood as a modification of the central
density in large amplitude motion leading to an excitation of the
breathing mode. The case of the GQR seems to be related to the extreme
collectivity of this state leading to a strong quadrupole response to
the quadrupole component of the non-linearities of the mean-field. We
also want to stress that, because of the perturbative nature of the
observed phenomenon, the possible introduction of four-phonon states
should not modify the above conclusions about two phonon states. Of
course, our findings imply that in order to get a correct three-phonon
spectrum one should further enlarge the space up to four-phonons. This
is a formidable task which is beyond the scopes of the present paper.

\begin {table} 
\caption { RPA one-phonon basis for $^{40}$Ca. For each state,
spin, parity, isospin, energy and percentage of the EWSR are
reported.  
In the following columns, $V_{M_1}$ stands for the matrix
element $<\nu|V|\nu~$$\otimes$$~M_1>$, where $\nu$ is the one phonon 
in  the $1^{st}$
column, the same for $V_{M_2}$ and $V_{Q_1}$. }
\label{ca40.rpa}
\begin{center}
\begin{tabular}{lccrccc}
\tiny Phonons  &\tiny $J^\pi~T$ & \tiny$E(MeV)$ & \tiny$\%EWSR$ & \tiny$V_{M_1}(MeV)$ & \tiny$V_{M_2}(MeV)$ & \tiny$V_{Q_1}(MeV)$\\ 
\hline
$ M_1  $&$ 0^+ ~0   $&$ 18.25  $&$ 30$&$-2.13 $&$-2.36 $&$- $\\
$ M_2  $&$ 0^+ ~0   $&$ 22.47  $&$ 54$&$-2.03 $&$-3.96 $&$- $\\ \hline
$ D_1  $&$ 1^- ~1   $&$ 17.78  $&$ 56$&$-1.38 $&$-2.12 $&$-1.25 $\\
$ D_2  $&$ 1^- ~1   $&$ 22.03  $&$ 10$&$-1.48 $&$-2.16 $&$+0.73 $\\ \hline
$ Q_1  $&$ 2^+ ~0   $&$ 16.91  $&$ 85$&$-1.36 $&$-2.49 $&$-0.36 $\\
$ Q_2  $&$ 2^+ ~1   $&$ 29.59  $&$ 26$&$-1.70 $&$-2.85 $&$-0.00 $\\ \hline
$ 3^-  $&$ 3^- ~0   $&$ ~4.94  $&$ 14$&$-1.74 $&$-2.60 $&$-0.07 $\\
$ O_1  $&$ 3^- ~0   $&$ ~9.71  $&$ ~5$&$-1.42 $&$-2.28 $&$-0.43 $\\
$ O_2  $&$ 3^- ~0   $&$ 31.33  $&$ 25$&$-1.69 $&$-2.72 $&$-0.31 $\\ 
\end{tabular}
\end{center}
\end{table}
\begin {table} 
\caption {Results for $^{40}$Ca. In the first column, the states
are labelled by their main component in the eigenvector and their
 unperturbed energy (in parentheses). In the
second column, the amplitude of the main component $c_0$. Then for
each total angular momentum J, we show the results of the calculation in
the basis up to 2 phonon states, the present results for the basis
extended to 3 phonon states, the corresponding first order
perturbation theory energy, and the second order one. The last two
columns contain the 2$^{nd}$ main component in the eigenstates and the
corresponding amplitude $c_1$. The subindex in the two-phonon
configurations denotes J. All energies are given in MeV.}
\label{ca40.res}
\begin{tabular}{lccrccccc}
\tiny Main & \tiny $c_0$ & \tiny $J^\pi$ & \tiny$\le2ph$ 
& \tiny$\le3ph$ & \tiny$1^{st}$ & \tiny$2^{nd}$
& \tiny$2^{nd}main$ & \tiny $c_1 $\\ 
\tiny component& $$ & $$ & $$ & $$ & \tiny$order$ & \tiny$order$ & \tiny component &   \\ 
\tableline
\tiny$3^-\!\!$$~\otimes$$\!\!~~3^- $&\tiny$-0.91  $&\tiny$ 0^+ $&\tiny$ 10.96 $&\tiny$ ~9.27 $&\tiny$
12.12 $&\tiny$ 9.20 $&$M_1 $&\tiny$
~0.21 $\\
$(~9.88)$&\tiny$ -0.96 $&\tiny$ 2^+ $&\tiny$ 10.63 $&\tiny$ ~8.89 $&\tiny$ 10.66 $&\tiny$ ~8.75
$&\tiny$(3^-)^2_2$$~\otimes$$\!\!~~M_2  $&\tiny$-0.21 $\\
&\tiny$ -0.96$&\tiny$ 4^+ $&\tiny$ ~9.85 $&\tiny$ ~8.10 $&\tiny$ ~9.86 $&\tiny$ ~7.96 $&\tiny$(3^-)^2_4
 $$~\otimes$$\!\!~~M_2  $&\tiny$-0.21 $\\
&\tiny$ -0.96$&\tiny$ 6^+ $&\tiny$ 10.88 $&\tiny$ ~9.12 $&\tiny$ 10.88 $&\tiny$ ~8.99 $&\tiny$(3^-)^2_6$
 $~\otimes$$\!\!~~M_2  $&\tiny$-0.21 $\\ \hline

$D_1\!\!$$~\otimes$$\!\!~~D_1 $&\tiny$-0.92  $&\tiny$ 0^+ $&\tiny$ 35.27 $&\tiny$ 33.71 $&\tiny$
35.25 $&\tiny$ 33.59 $&\tiny$(3^-)^2_0 $$~\otimes$$\!\!~~M_2 $&\tiny$
-0.22 $\\
$(35.56) $&\tiny$-0.96 $&\tiny$ 2^+ $&\tiny$ 35.10 $&\tiny$ 33.66 $&\tiny$ 35.06 $&\tiny$ 33.59
 $&\tiny$(D_1)^2_2$$~\otimes$$\!\!~~M_2  $&\tiny$-0.17 $\\ \hline

$D_1\!\!$$~\otimes$$\!\!~~Q_1 $&\tiny$ ~0.95 $&\tiny$ 1^- $&\tiny$ 34.83 $&\tiny$ 33.35 $&\tiny$
34.72 $&\tiny$ 33.24 $&\tiny$(M_2 \!\!$$~\otimes$$\!\!~~D_1)_1\!\!$$~\otimes$$\!\!~~Q_1 $&\tiny$
~0.19 $\\
$(34.69) $&\tiny$ 0.96$&\tiny$ 2^- $&\tiny$ 34.56 $&\tiny$ 33.22 $&\tiny$ 34.56 $&\tiny$ 33.16 $&\tiny$(M_2
 \!\!$$~\otimes$$\!\!~~D_1)_1\!\!$$~\otimes$$\!\!~~Q_1  $&\tiny$~0.19 $\\
 &\tiny$-0.96 $&\tiny$ 3^- $&\tiny$ 34.67 $&\tiny$ 33.13 $&\tiny$ 34.67 $&\tiny$ 33.02 $&\tiny$
(M_2 \!\!$$~\otimes$$\!\!~~D_1)_1\!\!$$~\otimes$$\!\!~~Q_1   $&\tiny$-0.19 $\\ \hline

$Q_1\!\!$$~\otimes$$\!\!~~Q_1 $&\tiny$ -0.87 $&\tiny$ 0^+ $&\tiny$ 33.88 $&\tiny$ 32.47 $&\tiny$
33.83 $&\tiny$ 32.27 $&\tiny$(Q_1 \!\!$$~\otimes$$\!\!~~3^-)_3\!\!$$~\otimes$$\!\!~~O_1 $&\tiny$
~0.32 $\\
$(33.82) $&\tiny$~0.84 $&\tiny$ 2^+ $&\tiny$ 33.82 $&\tiny$ 32.47 $&\tiny$ 33.82 $&\tiny$ 32.26 $&\tiny$(Q_1
 \!\!$$~\otimes$$\!\!~~3^-)_5\!\!$$~\otimes$$\!\!~~O_1 $&\tiny$-0.38 $\\
 &\tiny$ ~0.90$&\tiny$ 4^+ $&\tiny$ 34.02 $&\tiny$ 32.61 $&\tiny$ 34.02 $&\tiny$ 32.44 $&\tiny$(Q_1
 \!\!$$~\otimes$$\!\!~~3^-)_5\!\!$$~\otimes$$\!\!~~O_1 $&\tiny$ -0.32$\\ \hline

$M_2\!\!$$~\otimes$$\!\!~~D_1 $&\tiny$ -0.89 $&\tiny$ 1^- $&\tiny$40.26$&\tiny$ 38.14 $&\tiny$
40.05 $&\tiny$ 37.65 $&\tiny$(M_2)^2_0$$~\otimes$$\!\!~~D_1 $&\tiny$
~0.26 $\\
$(40.25) $& &\tiny$  $&\tiny$  $&\tiny$  $&\tiny$  $&\tiny$  $&\tiny$ $&\tiny$ $\\ \hline

$M_2\!\!$$~\otimes$$\!\!~~Q_1 $&\tiny$ -0.73 $&\tiny$ 2^+ $&\tiny$39.62$&\tiny$ 37.34 $&\tiny$
39.35 $&\tiny$ 36.80 $&\tiny$(O_1)^2_2 $$~\otimes$$\!\!~~M_1 $&\tiny$
~0.40 $\\ 
$(39.38) $& &\tiny$  $&\tiny$  $&\tiny$  $&\tiny$  $&\tiny$  $&\tiny$ $&\tiny$ $\\ \hline

$M_2\!\!$$~\otimes$$\!\!~~M_2 $&\tiny$ ~0.67 $&\tiny$ 0^+ $&\tiny$ 45.60 $&\tiny$ 42.76 $&\tiny$
44.87 $&\tiny$ 41.18 $&\tiny$(O_1)^2_0 $$~\otimes$$\!\!~~M_2 $&\tiny$
-0.55 $\\ 
$(44.94) $& &\tiny$  $&\tiny$  $&\tiny$  $&\tiny$  $&\tiny$  $&\tiny$ $&\tiny$ $\\ 

\end{tabular}

\end{table}

\begin {table} 
\caption { Same as table I for the nucleus $^{208}$Pb.}
\label{pb.rpa}
\begin{center}

\begin{tabular}{lccrccc}
\tiny Phonons & \tiny $J^\pi~T$ & \tiny$E(MeV)$ & \tiny$\%EWSR$ & \tiny$V_{M_1}(MeV)$ & \tiny$V_{M_2}(MeV)$ & \tiny$V_{Q_1}(MeV)$\\ 
\hline
$ M_1  $&$ 0^+ ~0   $&$ 13.61  $&$ 61$&$-1.87 $&$-0.92 $&$- $\\
$ M_2  $&$ 0^+ ~0   $&$ 15.02  $&$ 28$&$-1.32 $&$-1.16 $&$- $\\ \hline
$ D_1  $&$ 1^- ~1   $&$ 12.43  $&$ 63$&$-0.79 $&$-0.59 $&$-0.68 $\\
$ D_2  $&$ 1^- ~1   $&$ 16.66  $&$ 17$&$~0.00 $&$~0.00 $&$-0.64 $\\ \hline
$ 2^+  $&$ 2^+ ~0   $&$ ~5.54  $&$ 15$&$-0.11 $&$~0.07 $&$-1.18 $\\
$ Q_1  $&$ 2^+ ~0   $&$ 11.60  $&$ 76$&$-0.64 $&$-0.48 $&$-0.74 $\\
$ Q_2  $&$ 2^+ ~1   $&$ 21.81  $&$ 45$&$-0.86 $&$-0.63 $&$-0.55 $\\ \hline
$ 3^-  $&$ 3^- ~0   $&$ ~3.46  $&$ 21$&$-1.13 $&$-0.62 $&$-0.90 $\\
$ O    $&$ 3^- ~0   $&$ 21.30  $&$ 37$&$-0.99 $&$-0.74 $&$-0.42 $\\
\end{tabular}
\end{center}

\end{table}

\begin {table} 
\caption {Same as table II for the $^{208}$Pb nucleus.}
\label{pb.res}

\begin{tabular}{lccrccccc}
\tiny Main & \tiny $c_0$ & \tiny $J^\pi$ & \tiny$\le2ph$ 
& \tiny$\le3ph$ & \tiny$1^{st}$ & \tiny$2^{nd}$
& \tiny$2^{nd}main$ & \tiny $c_1 $\\ 
\tiny component& $$ & $$ & $$ & $$ & \tiny$order$ & \tiny$order$ & \tiny component &   \\ 
\tableline
\tiny$3^-\!\!$$~\otimes$$\!\!~~3^- $&\tiny$ -0.95 $&\tiny$ 0^+ $&\tiny$ ~7.88 $&\tiny$ ~6.96 $&\tiny$
8.06 $&\tiny$ 6.90 $&\tiny$(3^-)^2_0$$~\otimes$$\!\!~~2^+ $&\tiny$
-0.17 $\\
$(~6.93)$&\tiny$-0.92$ &\tiny$ 2^+ $&\tiny$ ~7.31 $&\tiny$ ~6.57 $&\tiny$ ~7.33 $&\tiny$ ~6.52 $&\tiny$2^+
$&\tiny$-0.28 $\\
&\tiny$-0.98$ &\tiny$ 4^+ $&\tiny$ ~7.16  $&\tiny$ ~6.55 $&\tiny$ ~7.16 $&\tiny$ ~6.51 $&\tiny$(3^-)^2_4
 $$~\otimes$$\!\!~~M_1  $&\tiny$-0.15 $\\
&\tiny$~0.97$ &\tiny$ 6^+ $&\tiny$~7.43  $&\tiny$ ~6.63 $&\tiny$ ~7.44 $&\tiny$ ~6.56 $&\tiny$(3^-)^2_6
 $$~\otimes$$\!\!~~M_1  $&\tiny$~0.15 $\\ \hline

$3^-\!\!$$~\otimes$$\!\!~~2^+ $&\tiny$ -0.94 $&\tiny$ 1^- $&\tiny$ ~9.20 $&\tiny$ ~8.26 $&\tiny$
9.21 $&\tiny$ 8.02 $&\tiny$(2^+)^2_2 $$~\otimes$$\!\!~~3^- $&\tiny$
-0.23 $\\
$(~9.01)$&\tiny$~0.97 $&\tiny$ 2^- $&\tiny$ ~9.12 $&\tiny$ ~8.54 $&\tiny$ ~9.12 $&\tiny$ ~8.50 $&\tiny$(2^+)^2_2
$$~\otimes$$\!\!~~3^-  $&\tiny$~0.17 $\\
&\tiny$ ~0.96$&\tiny$ 3^- $&\tiny$ ~9.17 $&\tiny$ ~8.70 $&\tiny$ ~9.12 $&\tiny$ ~8.56 $&\tiny$(3^-)^3_2
  $&\tiny$-0.17 $\\
&\tiny$ ~0.96$&\tiny$ 4^- $&\tiny$ ~9.07 $&\tiny$ ~8.61 $&\tiny$ ~9.07 $&\tiny$ ~8.45 $&\tiny$(3^-)^3
  $&\tiny$~0.19 $\\
&\tiny$ -0.96$&\tiny$ 5^- $&\tiny$ ~9.06 $&\tiny$ ~8.33 $&\tiny$ ~9.06 $&\tiny$ ~8.16 $&\tiny$(2^+)^2_2 
$$~\otimes$$\!\!~~3^-  $&\tiny$-0.18 $\\ \hline

$2^+\!\!$$~\otimes$$\!\!~~2^+ $&\tiny$ ~0.92 $&\tiny$ 0^+ $&\tiny$ ~11.23 $&\tiny$ ~9.88 $&\tiny$
11.24 $&\tiny$ 9.46 $&\tiny$(2^+)^3  $&\tiny$
-0.31 $\\
$(11.09)$&\tiny$-0.94 $&\tiny$ 2^+ $&\tiny$ 11.27 $&\tiny$ ~10.78 $&\tiny$ 11.12 $&\tiny$ 10.61 $&\tiny$(3^-)^2_2
 $$~\otimes$$\!\!~~2^+  $&\tiny$~0.24 $\\
&\tiny$~0.94 $&\tiny$ 4^+ $&\tiny$ 11.25 $&\tiny$ 10.39 $&\tiny$ 11.25 $&\tiny$ 10.13 $&\tiny$(2^+)^3
 $&\tiny$~0.24 $\\ \hline

$D_1\!\!$$~\otimes$$\!\!~~D_1 $&\tiny$ ~0.97 $&\tiny$ 0^+ $&\tiny$ 24.91 $&\tiny$ 24.42 $&\tiny$
24.90 $&\tiny$ 24.40 $&\tiny$(D_1)^2_0 $$~\otimes$$\!\!~~M_1 $&\tiny$
~0.11 $\\
$(24.87)$ &\tiny$~0.96 $&\tiny$ 2^+ $&\tiny$ 24.68 $&\tiny$ 24.29 $&\tiny$ 24.68 $&\tiny$ 24.27
 $&\tiny$3^-\!\!$$~\otimes$$\!\!~~O  $&\tiny$~0.19 $\\ \hline

$D_1\!\!$$~\otimes$$\!\!~~Q_1 $&\tiny$-0.96  $&\tiny$ 1^- $&\tiny$ 24.07 $&\tiny$ 23.73 $&\tiny$
24.02 $&\tiny$ 23.71 $&\tiny$(3^-)_2 $$~\otimes$$\!\!~~D_1 $&\tiny$
~0.17 $\\
$(24.03)$&\tiny$~0.98 $&\tiny$ 2^- $&\tiny$ 23.97  $&\tiny$ 23.82 $&\tiny$ 23.97 $&\tiny$ 23.80 $&\tiny$(3^-)^2_2
$$~\otimes$$\!\!~~D_1  $&\tiny$-0.16 $\\
 &\tiny$~0.96 $&\tiny$ 3^- $&\tiny$ 24.03 $&\tiny$ 23.74 $&\tiny$ 24.03 $&\tiny$ 23.71 $&\tiny$
(2^+)^2_2 $$~\otimes$$\!\!~~D_1   $&\tiny$~0.18 $\\ \hline

$Q_1\!\!$$~\otimes$$\!\!~~Q_1 $&\tiny$ -0.94 $&\tiny$ 0^+ $&\tiny$ 23.20 $&\tiny$ 22.92 $&\tiny$
23.20 $&\tiny$ 22.86 $&\tiny$(3^-)^2_2$$~\otimes$$\!\!~~Q_1 $&\tiny$
-0.24 $\\
$(23.20)$&\tiny$~0.95 $&\tiny$ 2^+ $&\tiny$ 23.23 $&\tiny$ 23.17 $&\tiny$ 23.18 $&\tiny$ 23.14 $&\tiny$(3^-)^2_2
 $$~\otimes$$\!\!~~Q_1 $&\tiny$~0.22 $\\
 &\tiny$-0.95 $&\tiny$ 4^+ $&\tiny$ 23.26 $&\tiny$ 23.10 $&\tiny$ 23.26 $&\tiny$ 23.07 $&\tiny$(3^-)^2_2
$$~\otimes$$\!\!~~Q_1  $&\tiny$ -0.22$\\ \hline

$M_1\!\!$$~\otimes$$\!\!~~D_1 $&\tiny$-0.94  $&\tiny$ 1^- $&\tiny$ 26.05 $&\tiny$ 25.35 $&\tiny$
26.02 $&\tiny$ 25.28 $&\tiny$(M_1)^2_0 $$~\otimes$$\!\!~~D_1 $&\tiny$
-0.20 $\\
$(26.05) $& &\tiny$  $&\tiny$  $&\tiny$  $&\tiny$  $&\tiny$  $&\tiny$ $&\tiny$ $\\ \hline

$M_1\!\!$$~\otimes$$\!\!~~Q_1 $&\tiny$-0.92  $&\tiny$ 2^+ $&\tiny$ 25.25 $&\tiny$ 24.77 $&\tiny$
25.22 $&\tiny$ 24.66 $&\tiny$(3^-)_2^2 $$~\otimes$$\!\!~~M_1 $&\tiny$
-0.20 $\\
$(25.21) $& &\tiny$  $&\tiny$  $&\tiny$  $&\tiny$  $&\tiny$  $&\tiny$ $&\tiny$ $\\ \hline

$M_1\!\!$$~\otimes$$\!\!~~M_1 $&\tiny$~0.74  $&\tiny$ 0^+ $&\tiny$ 27.52 $&\tiny$ 26.23 $&\tiny$
27.28 $&\tiny$ 25.95 $&\tiny$(2^+)^2_0 $$~\otimes$$\!\!~~M_2 $&\tiny$
~0.54 $\\
$(27.22) $& &\tiny$  $&\tiny$  $&\tiny$  $&\tiny$  $&\tiny$  $&\tiny$ $&\tiny$ $\\ 

\end{tabular}

\end{table}
\end{document}